# Challenges and Opportunities in Searching for Rashba-Dresselhaus Materials for Efficient Spin-Charge Interconversion at Room Temperature

Zixu Wang[1], Zhizhong Chen[1], Rui Xu[2], Hanyu Zhu[2,*], Ravishankar Sundararaman[1,3,*], Jian Shi[1,3,*]

[1]Department of Materials Science and Engineering, Rensselaer Polytechnic Institute, Troy, NY 12180, USA.
[2]Department of Materials Science and NanoEngineering, Rice University, Houston, TX 77005, USA
[3]Department of Physics, Applied Physics, and Astronomy, Rensselaer Polytechnic Institute, Troy, NY 12180, USA.

Correspondence:[*] H. Z.: hanyu.zhu@rice.edu; R.S.: sundar@rpi.edu; J.S.: shij4@rpi.edu

**Abstract:**

Spintronic logic devices require efficient spin-charge interconversion: converting charge current to spin current and spin current to charge current. In spin-orbit materials that are regarded as the most promising candidate for spintronic logic devices, one mechanism that is responsible for spin-charge interconversion is Edelstein and inverse Edelstein effects based on spin-momentum locking in materials with Rashba-type spin-orbit coupling. Over last decade, there has been rapid progresses for increasing interconversion efficiencies due to the Edelstein effect in a few Rashba-Dresselhaus materials and topological insulators, making Rashba spin-momentum locking a promising technological solution for spin-orbit logic devices. However, despite the rapid progress that leads to high spin-charge interconversion efficiency at cryogenic temperatures, the room-temperature efficiency needed for technological applications is still low. This paper presents our understanding on the challenges and opportunities in searching for Rashba-Dresselhaus materials for efficient spin-charge interconversion at room temperature by focusing on materials properties such as Rashba coefficients, momentum relaxation times, spin-momentum locking relations and electrical conductivities.

**Introduction:**

In addressing the issues of scaling and energy efficiency CMOS technology is facing, a promising proposal is spintronics which harnesses the spin degrees of freedom for logic or memory application. Compared to charge-based operation in CMOS whose operation efficiency and voltage is limited, spin-based spintronic devices have been evaluated with potential advantages on low energy operation, high energy efficiency, high logic density and fast switching[1].

There has been a plethora of designs on using spin degrees of freedom for logic operation. For example, the famous Datta-Das spin field effect transistor[2] and its modified version (e.g. using charge-spin conversion materials replacing ferromagnetic electrodes) utilizes the Rashba spin-orbit effective magnetic field to precess electron spin, in which the spin precession angle can be controlled by externally applied electric field. Another design that has been thoroughly evaluated is Intel's recently proposed magneto-electric spin-orbit (MESO) technology[1] and its variants (e.g. all spin-orbit logic)[3]. It suggests that by harnessing the merits of efficient spin-charge conversion in materials with strong spin-orbit coupling one can reduce the power use of logic devices by more than one order of magnitude (10~30 ×) with enhanced logic density (5 ×) compared to CMOS (with similar or better computing performance in terms of switching speed and the number of logic operations per unit area and unit time). In many of these designs, spin-charge interconversion is a central function. The high efficiency at room temperature of spin-charge interconversion is a prerequisite for practical applications.

Among many mechanisms, spin-charge interconversion through spin-orbit coupling has been mostly studied due to their promises in terms of conversion efficiency and scalability[4-6]. There are two major conversion mechanisms: (1) spin Hall effect that converts charge current to spin current; inverse spin Hall effect that converts spin current to charge current; (2) Edelstein effect that converts charge current to spin current; inverse Edelstein effect that converts spin current to charge current. The underlying physics responsible for the intrinsic spin Hall effect (a 3D effect) and inverse spin Hall effect is based on Berry curvature which is spin-dependent in spin-orbit coupled materials. The charge-spin conversion efficiency in spin Hall materials is defined as spin Hall angle $\theta_{SH}$ (here $\theta_{SH} \equiv \frac{\sigma_{xy}^S}{\sigma_{xx}^c}$, $\sigma_{xy}^S$ spin Hall conductivity, $\sigma_{xx}^c$ conductivity) which is usually much smaller than one. Typical top-performed spin Hall materials include elemental metals such as $\beta$-W, $\beta$-Ta and Pt (with spin Hall angle as 0.4, 0.15 and 0.08, respectively, at room temperature)[4].

Edelstein effect and inverse Edelstein effect are found in both Rashba materials and the surface states of topological insulators both of which have Rashba-type spin-momentum locking[5, 7-12]. For both Edelstein effect and inverse Edelstein effect (essentially a 2D effect), the spin-charge interconversion is achieved by charge/spin current-induced shift of Fermi contour that is accompanied by the generation of pure spin or pure charge current (Fig. 1). It has been found that Edelstein effect and inverse Edelstein effect may deliver much higher conversion efficiency than spin Hall effect. For example, in 2016, A Fert et al[12] have discovered that at room temperature, through Edelstein effect, the surface states of topological insulator $\alpha$-Sn can have a spin-to-charge conversion length $\lambda_{IEE}$ at 2.1 nm (an equivalent spin Hall angle of 1.6; here $\lambda_{IEE} \equiv \frac{j_c^{2D}}{j_s^{3D}}$, $j_s^{3D}$ spin current density, $j_c^{2D}$ charge current density; the charge-to-spin conversion efficiency is characterized by $q_{EE} \equiv \frac{j_s^{3D}}{j_c^{2D}}$ but a common figure of merit for charge-to-spin conversion in Rashba-Dresselhauss materials or topological insulators is the spin-orbit-torque ratio $\theta_{||} = \frac{\sigma_S}{\sigma}$ where $\sigma_S$ is spin Hall conductivity and $\sigma$ charge conductivity). Since the first experimental demonstration of the Edelstein effect, researchers have shown steady progress in improving the spin-charge interconversion efficiency in both Rashba materials and topological insulators. Examples[1, 13-30] include Rashba interfaces LaAlO$_3$/SrTiO$_3$, Bi/Ag/Al$_2$O$_3$, Al/SrTiO$_3$, 2D materials, and topological insulators $\alpha$-Sn, Bi$_2$Se$_3$, BiSe, and (Bi$_{1-x}$Sb$_x$)$_2$Te$_3$.

**<u>Challenges:</u>**

For practical spintronic device application (e.g. MESO) using Edelstein effect, spin-charge conversion efficiency, in terms of conversion length $\lambda_{IEE}$ for example, above 5 nm, is suggested[1]. Currently studied Rashba materials or topological insulators unfortunately hardly deliver such a high efficiency at room temperature. At cryogenic temperature, some model systems such as SrTiO$_3$-based 2D electron gas can deliver a conversion length at 20 nm (e.g. at 15 K by M Bibes et al[31] in 2019) but it drops to 0.5 nm at room temperature. So what are the limiting factors on improving room temperature conversion efficiency?

With simple circular Fermi contours with typical 2D Rashba helical spin-momentum locking shown in Fig. 1a as an example, when an external electric field is applied along the -$x$ direction, we expect a shift of Fermi contours (e.g. conduction band) towards +$x$ direction with its net momentum proportion to the strength of the external electric field and the momentum relaxation time. Such a shift leads to the accumulation of non-zero spin density along +$y$ direction (Fig. 1b) since the inner Fermi circle (with opposite spin direction) cannot be perfectly balanced by the outer circle. This is the Edelstein effect (or Rashba-Edelstein effect) that converts charge to spin. When a spin current is injected into the system (e.g. the direction of spin polarization is along $+y$), the imbalanced spin density reciprocally produces a net charge momentum (e.g. along $+x$ direction) (Fig. 1c and d). This is the inverse Edelstein effect (or inverse Rashba-Edelstein effect). When dealing with the surface states of topological insulators which have similar spin-momentum locking relation as either inner or outer Fermi circles of a typical 2D Rashba system, one only considers one Fermi circle which also removes the canceling effect from the other Fermi circle seen in Rashba system.

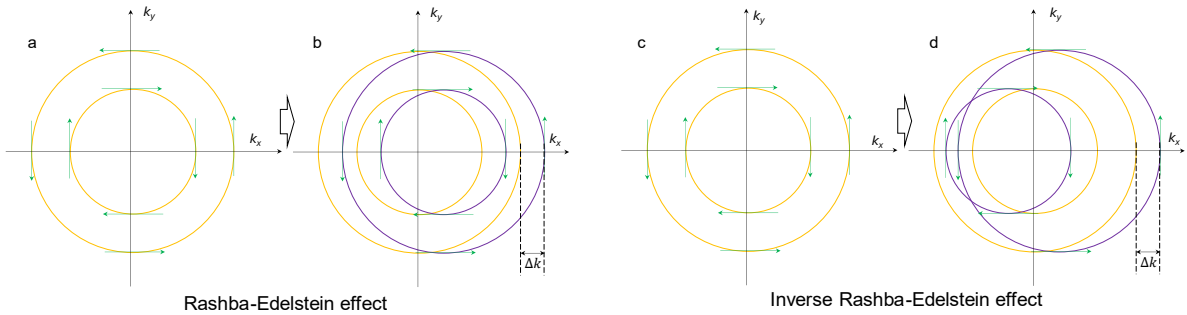

*Fig. 1 Spin-charge interconversion by Rashba-Edelstein effect and Inverse Rashba-Edelstein effect.*

Analytically, with circular Fermi disks, for Rashba systems with helical spin-momentum locking (Fig. 1), we can arrive the following relation[32]: $\lambda_{IEE} = \alpha_R \tau_m/\hbar$, $\alpha_R$: Rashba coefficient; $\tau_m$: momentum relaxation time; for the surface states of topological insulator, we have $\lambda_{IEE} = v_F \tau_m$, $v_F$ Fermi velocity. From these equations, one can simply arrive at following arguments: for enhancing $\lambda_{IEE}$, one needs to search for materials with large Rashba coefficients, momentum relaxation time, and/or Fermi velocity[33]. It is noted that spin transmission time at the interface also influences the spin-charge conversion efficiency[34]. For charge-to-spin conversion figure of merit $\theta_{||}$, it has been found that it also scales with the Rashba coefficient in Rashba materials[34]. These are very stringent requirements. With topological insulators as one example, to harness its surface state, the existence of room temperature band gap and nontrivial topology for the bulk phase of topological insulators is required, which itself is unfortunately a long-term technical challenge researchers have been working on, not to mention the engineering of the Fermi velocity and momentum relaxation time. For Rashba-Dresselhaus materials, the lift of requirements on these topological properties helps widen the materials space. However, there are not many Rashba-Dresselhaus materials reported with consistent experimental data (often due to the use of different measurement conditions) or computational predictions with $\lambda_{IEE}$ or $\theta_{||}$. And it is unclear how well one can use such a simplified equation to guide the materials design.

When considering real materials systems, the circular Fermi disk picture with perpendicular spin-momentum relation could be no longer valid for Rashba materials or even the surface states of topological insulators (e.g. the warping[35] of spin-momentum locking in the surface states of topological $Bi_2Se_3$). With a non-circular Fermi disk and other spin-momentum locking relation, one cannot simply adopt the above analytic approximation to estimate conversion length since its impact on $\lambda_{IEE}$ and $\theta_{||}$ can be enormous (e.g. orders of magnitude difference in $SrTiO_3$-based system[31]).

Thus, a fundamental question arises: what are the materials design space that could allow us to receive large $\lambda_{IEE}$ and $\theta_{||}$ at room temperature?

**Opportunities:**

When considering Rashba coefficients for both $\lambda_{IEE}$ and $\theta_{||}$, a natural choice for candidate materials would be those with heavy elements such as Bi, Ag, Pb and so on which have strong spin-orbit coupling (Fig. 2). Data on Fig. 2 are collected from references[36-45], from which it is seen that materials with large Rashba coefficients usually contain heavy elements such as Bi, Pb and I. The existence of Rashba effect requires inversion symmetry breaking (either bulk inversion symmetry breaking or interface-induced inversion symmetry breaking). It is noted that the Rashba effect[46] from organic-inorganic hybrid perovskite $CH_3NH_3PbI_3$ is likely from its surface inversion symmetry breaking (experimentally it is found that local structural fluctuation is averaged out at the bulk form). For a model 2D Rashba system where the Rashba field comes from the externally applied field, the Rashba coefficients also scale with the strength of the electric field. For Rashba heterostructures where Rashba effect is from the interfacial inversion symmetry breaking, one considers the effective field strength in the interface. Thus, one should not only explore materials containing heavy elements but also search for heterostructures that may have large differences in work functions for metals or electron affinities for semiconductors. Strain could be also used as a knob for lifting inversion symmetry[47] to enable Rashba effect and/or tuning the gradient of crystal potential[48, 49] for further escalating Rashba coefficients. Further, it is also found that surface reconstruction in perovskite halides could also lead to static Rashba effect[50]. Another approach for engineering Rashba effect is through chemical composition gradient which breaks inversion symmetry. When heavy elements are involved, room temperature Rashba effect may be observed[51]. Additionally, large Rashba coefficients also suggest large Rashba splitting energy $E_R$ (Fig. 3). For applications such as spin field effect transistor operating at room temperature, $E_R$ larger than 26 meV is required.

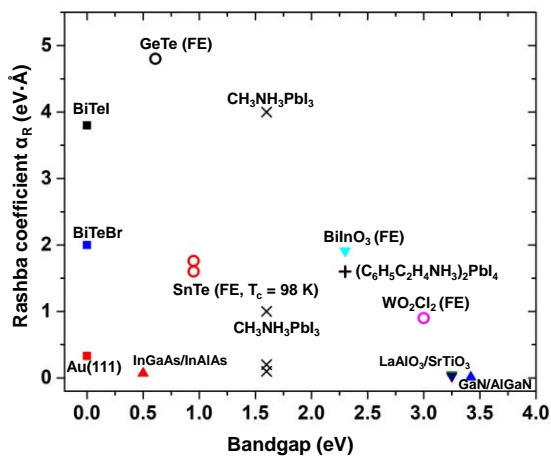

*Fig. 2 Rashba coefficients and band gaps for a list of materials.*

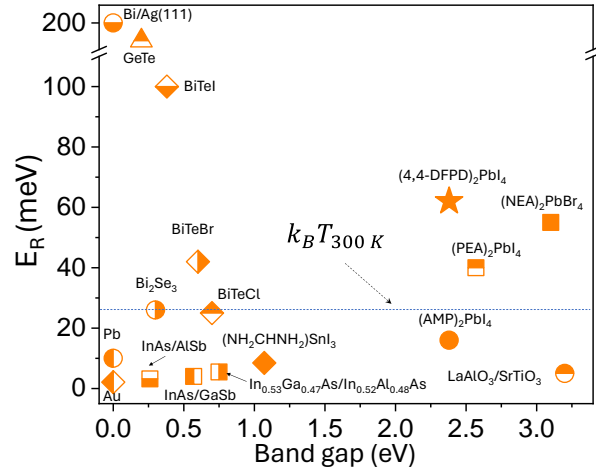

*Fig. 3 Rashba splitting energy and band gaps for a list of materials. Adapted from ref. 43 with permission.*

For momentum relaxation time that at least has strong influence on $\lambda_{IEE}$, it is mainly determined by electron-phonon coupling and defects scattering. For electron-phonon coupling in the Rashba system, one thus needs to consider the phonon occupancy and electron-phonon coupling strength. Polar systems tend to have stronger electron-phonon coupling strength and stronger temperature-dependent conversion efficiency, but one can search for materials systems that have low phonon occupancy at room temperature. Defects scattering can be suppressed by reducing physical defects at the interface or through modulation doping.

The understanding of the roles of spin texture or spin-momentum locking relation on spin-charge interconversion is limited. Most studies assume simple circular Fermi contours for both Rashba interfaces (Fig. 4a and b) or the surface states of topological insulators. These assumptions are crude that can lead to less accurate estimation on the conversion performance. Further, the spin-canceling Fermi contours in the 2D Rashba system are hardly the most efficient configurations for charge-spin conversion (Fig. 4c and f). Even for the surface states of topological insulators which have one Rashba-like Fermi contour that does not have a canceling counterpart (Fig. 4i), whether such spin-momentum locking relation is the optimized configuration remains elusive (not to mention that the real spin-momentum locking could be warped from 2D-like to 3D-like shown in Fig. 4h or complicated by the presence of a surface Rashba splitting shown in Fig. 4k, l and m[7]). It has been experimentally[31] observed that by tuning the position of Fermi level one does not only see orders of magnitude change of conversion efficiency but also sign reversal assumed due to the change of spin texture.

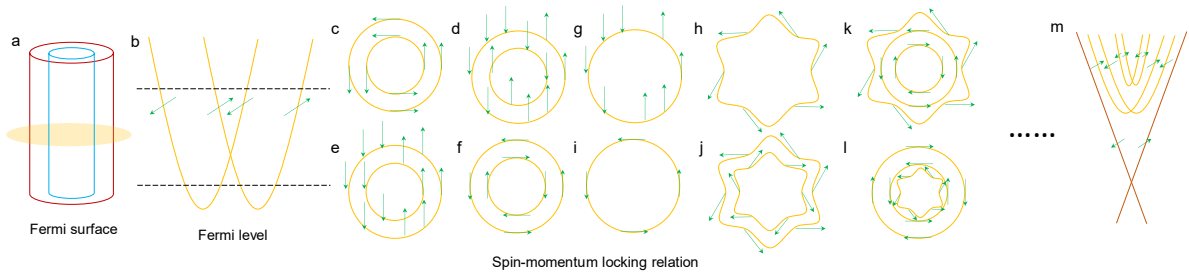

*Fig. 4 The vast configuration space of spin textures (including persistent spin helix-type) should be explored for designing spin-charge interconversion.*

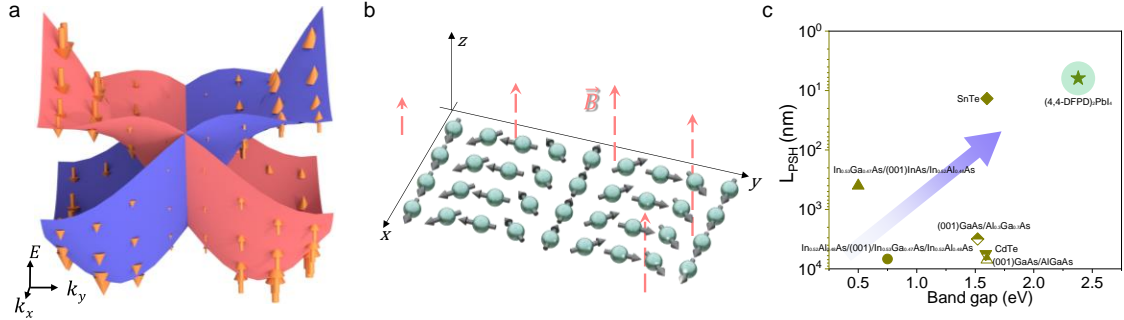

*Fig. 5 Persistent spin helix. (a) spin-orbit field in a mm2 crystal with quantum well structure. (b) a schematic drawing of persistent spin helix under a spin-orbit field. (c) a list of persistent spin helix materials with varied wavelength. Adapted from ref. 43 with permission.*

In fact, a simple calculation based on Boltzman transport, using the persistent spin helix-like spin texture (Fig. 4d and e) which can be found in the balanced Rashba-Dresselhaus [100] system or Dresselhaus [110] system rather than the classical 2D helical spin-momentum locking structure, suggests that one can increase $q_{EE}$ by a factor of $\frac{4}{\pi}$ when spin transmission time, Fermi velocity, and Rashba coefficients remain the same (e.g. Fig 4d versus Fig. 4f, or Fig. 4g versus 4i). Experimentally, such a type of spin texture has not been tested for evaluating the interconversion efficiency.

Previously, persistent spin helix was only demonstrated in 2D quantum well III-V materials at cryogenic temperature[52]. Recently Zhang et al discovered the existence of such states in a layered quantum well material at room temperature[43, 53], suggesting a possibility on harnessing persistent spin helix-type spin texture for room-temperature spin-charge interconversion (Fig. 5). Fig. 5b lists several materials or heterostructures hosting persistent spin helix states with varied wavelengths. From symmetry perspectives[54], a series of van der Waals or quasi van der Waals ferroelectric crystals carrying in-plane polarizations and quasi square basal plane structure could harbor persistent spin helix states[43, 53, 55].

For the electrical conductivity of the spintronic materials, the requirements in different types of spintronic devices are different. For example, in MESO device and its variants[1, 3], the target electrical resistivity is expected to be larger than $10\ m\Omega \cdot cm$. Electrical conductivity is a function of Fermi level, doping, momentum relaxation time, and Fermi contours. Shift of Fermi level through either chemical or electrostatic doping could lead to varied spin texture or momentum relaxation time which further impacts the spin-charge interconversion.

**<u>Outlook:</u>**

In logic spintronic devices, efficient spin-charge interconversion has been pursued for decades. In practice, one often has to lower the operation temperature to liquid nitrogen or helium regime. Operating spintronic devices at cryogenic temperature limits their technological applications. Finding the right materials that could operate at room temperature, is aligned with the need for device miniaturization, and provide high spin-charge interconversion efficiency, would help meet the requirements of high efficiency, fast speed, and scaling for spin-based computing. Enriched basic understanding of relation among spin-orbit coupling, symmetry and spin dynamics could also help shed lights on the design of novel spin bits or quantum bits for future computing.

**Acknowledgements:**
R.S. acknowledges support from the Department of Energy under grant No. DE-SC0023301. H.Z. acknowledge Welch Foundation C-2128 for support. J.S. acknowledges the support from the National Science Foundation under award No. 2031692 and 2314614, and NYSTAR under contract C180117.

**Data and code availability:**
Data and code are available from the corresponding authors upon reasonable request.

**Conflict of interest:**
The authors declare no competing financial interest.